\documentclass[10pt,journal,compsoc]{IEEEtran}\usepackage{titling}
\usepackage{graphicx}\usepackage{subfigure}\usepackage{amssymb,amsmath}\usepackage[sort&compress]{natbib}\parskip=1mm
\usepackage[mediumspace,mediumqspace,Grey,squaren]{SIunits}
\usepackage{rotating}\usepackage{booktabs}
\bibpunct[,]{[}{]}{,}{n}{,}{,}
\usepackage{epstopdf}
\usepackage{abstract}\renewcommand{\abstractname}{}    

\def\thesection{\arabic{section}.}
\def\thesubsection{\thesection\arabic{subsection}}
 \makeatletter
  \renewcommand\@seccntformat[1]{%
    \@ifundefined{#1@cntformat}%
    {\csname the#1\endcsname\quad}%
    {\csname #1@cntformat\endcsname}}
  \newcommand\section@cntformat{\thesection.\quad}
  \newcommand\subsubsection@cntformat{\thesubsubsection.\quad}
\makeatother
 
\pretitle{\begin{flushleft}\LARGE\sffamily}
\posttitle{\par\end{flushleft}\vskip 0.5em}
\predate{\begin{flushleft}\large\scshape}
\postdate{\par\end{flushleft}}
\date{}
\begin{document}

\title{\bf Data on Displacements caused by the Growth of Bainite in Steels}

\twocolumn[
  \begin{@twocolumnfalse}\maketitle
    \begin{abstract}
\noindent{\bf \large H. K. D. H Bhadeshia} \\ \\
\noindent{\em Materials Science and Metallurgy, University of Cambridge, U. K., {\rm hkdb@cam.ac.uk}} \\ \\
 
\noindent One of the important characteristics of solid-state phase transformations in steels is the choreography of atoms as they traverse the frontier between the parent and the product phases. Some transformations involve a chaotic motion of atoms consistent with long-range diffusion, and hence are closer to equilibrium than those that where a  disciplined transfer occurs. Since the pattern of atoms changes during transformation, a disciplined motion of atoms necessarily leads to a change in the shape of the transformed region, and like any deformation, such changes cause strains in the surrounding material.  These displacive transformations are therefore strain dominated, with the morphology, chemical composition and thermodynamic framework sensitive to the strain energy due to the shape change. Here we consider the published data that have been accumulated on the displacements associated with `bainite', a phase transformation product in steels that forms the basis of the world's first bulk nanocrystalline metal. \end{abstract}
  \end{@twocolumnfalse}\vskip 0.5cm
]

\section{\bf Introduction}

There are two essential kinds of solid-state phase transformations in metals and alloys. Reconstructive transformations involve the uncoordinated motion of atoms, with diffusion occurring to minimise the strain energy of transformation, and in the case of alloys, to facilitate the partitioning of  solutes between the parent and product phases until the chemical potential of each component becomes uniform across the phases. 

The second kind involves a homogeneous deformation of the parent structure into that of the product;  atoms on the substitutional lattice do not diffuse during transformation. The deformation causes a change in the shape of the transformed region, a change  that can be measured and related to the atomic mechanism of transformation \cite{Christian:1997,Bhadeshia:bainite}.  Transformations like these involve macroscopic displacements and hence are labelled `displacive'. 

There is additional complexity in the case of steels, where the crystal structure is described in terms of the iron and substitutional-solute atoms,  whereas the carbon that resides in interstices  between the large atoms and in displacive transformations is chaperoned into the new lattice. On the other hand, the greater mobility of the carbon atoms means that they can partition while the substitutional lattice is displaced \cite{Christian:1962}; but their diffusion has no consequence on the shape deformation accompanying the crystal structure change \cite{Christian:1962}. 

\begin{figure}[htdp]
\centering
\subfigure(a){
\includegraphics[width=0.15\textwidth]{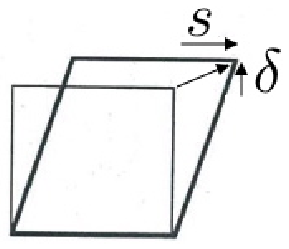}}
\subfigure(b){
\includegraphics[width=0.2\textwidth]{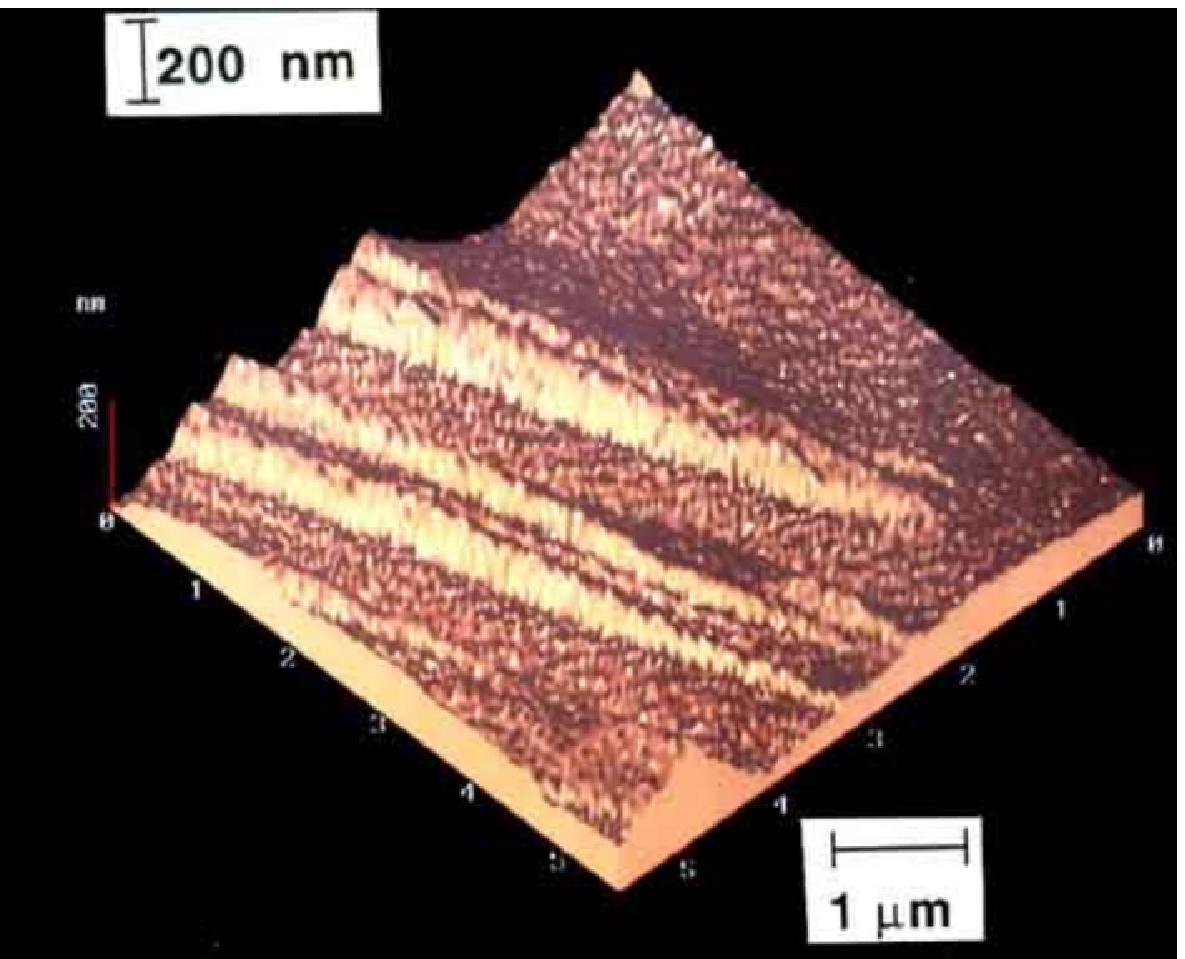}}
\caption{(a) Schematic illustration of a general invariant-plane strain defined by the cube of unit height changing into a parallelepiped. $s$ and $\delta$ represent the shear and dilatational strains respectively. The combination of $s$ and $\delta$ leads to the displacement $m$ that is not parallel to the habit plane (the horizontal invariant plane). (b) Actual shape change due to individual platelets of bainite \cite{Swallow:1996}.}
\label{fig:displacements}
\end{figure}

Throughout this paper, the parent phase is austenite, with a cubic close-packed crystal structure, and the product phase is bainitic ferrite, whose structure is conventionally regarded as body-centred cubic but can also have lower symmetry depending on the carbon that is present in solid solution \cite{Jang:2012b,Smith:2013,Bhadeshia:2013d}. Bainite is a displacive transformation and one which is of immense technological importance  \cite{Pickering:1967,Caballero:strong,Bhadeshia:2013}. All of the characteristics that make bainite so useful rely on the fact that it is a strain dominated transformation. THus, the  morphology and size of the bainite plates depend on the minimisation of strain energy due to the shape deformation. The  shape deformation is an invariant-plane strain, i.e., one which leaves a plane macroscopically unrotated and undistorted \cite{Christian:2003b,Christian:2003}; the invariant-plane is also the one on which the bainite plates lengthen. In detail, the shape deformation consists of a large shear on the invariant-plane and a small dilatation normal to that plane, as illustrated schematically on Fig.~\ref{fig:displacements}, which also  shows an actual image of the displacements produced when a single crystal of austenite is polished flat and allowed to transform into plates of bainite. This deformation creates a large dislocation density in the austenite, that in turn opposes the motion of the transformation interface \cite{Bhadeshia:1979}, which mechanically stabilises the austenite \cite{Chatterjee:2006} and hence stops the bainite platelets from coarsening.

The purpose of this paper is to assemble the data on measurements of the shape deformation of bainite, and to assess their value. We begin first with a description of the consequences of the shape deformation so that the measurements can be placed in context. As will be seen later, the measurements concerned are difficult and have to be conducted at a variety of resolutions.

\section{\bf Consequences of Shape Deformation}
Why is the shape deformation important in the theory and practice of solid-state phase transformations? Like any deformation that occurs within a metal, the material surrounding the individual plate is required to accommodate the displacements. In other words, compatibility requires that there are distortions in the matrix around each plate of bainite. For an elastically accommodated,  isolated plate in the form of an oblate spheroid with length $r$ much greater than the thickness $c$, located within elastically isotropic austenite, the strain energy per unit volume is given by \cite{Christian:1958}:
 \begin{equation}G_{\rm strain}= \frac{c}{r}\frac{\mu}{1-\nu}\biggl[\frac{\pi}{4}\delta^2 + \frac{\pi}{8}{(2-\nu)}s^2\biggr] \approx \frac{c}{r}\mu(s^2+\delta^2) \label{eqn:christian}
\end{equation}
where $\mu$ and $\nu$ are the respective shear modulus and Poisson's ratio of austenite, and $s$ and $\delta$ are respectively the shear and dilatational strains parallel and normal to the habit plane. 
For bainite, this strain energy comes to about 400\,J\,mol$^{-1}$ \cite{Bhadeshia:1980}, which is  large in the context of the chemical free energy change accompanying transformation to say allotriomorphic ferrite or pearlite. $G_{\rm strain}$ therefore dominates the bainite reaction, and is the sole reason why the product is in the form of thin plates ($c\ll r)$ because a small aspect ratio  minimises the strain energy. This of course is the same reason why martensite and mechanical twins form as thin plates.

Bainite forms at temperatures where the parent austenite is not strong. As a result, the shape deformation during transformation causes plastic relaxation in the adjacent austenite. The dislocation debris thus created in the austenite then opposes further transformation by a phenomenon known as mechanical stabilisation \cite{Strife:1977,Shipway:1995,Yang:1995b,Singh:1996,Yang:1996,Larn:2000}, and brings the growth of a plate to a halt before it impinges with hard obstacles \cite{Chatterjee:2006}. This has technological consequences because it leads to a dramatic refinement of the structure, and the increase in dislocation density contributes to strength. 

Disciplined movements of atoms cannot in general be sustained across crystal boundaries, so unlike diffusional transformations, plates of bainite are restricted to grow within the grains in which they nucleate.

\section{\bf The Data}

\subsection{\bf Qualitative data}
\begin{figure}[htdp]
\centering
\subfigure(a){
\includegraphics[width=0.45\textwidth]{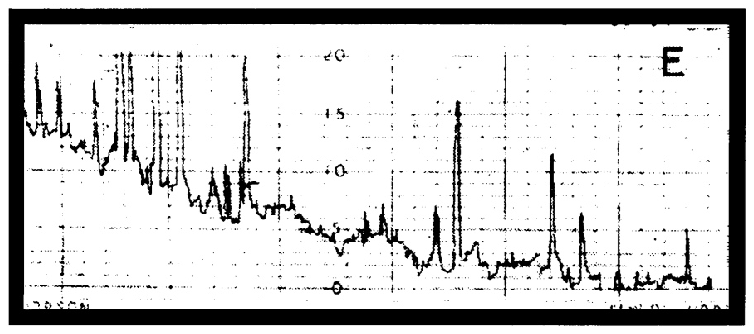}}\\
\subfigure(b){
\includegraphics[width=0.40\textwidth]{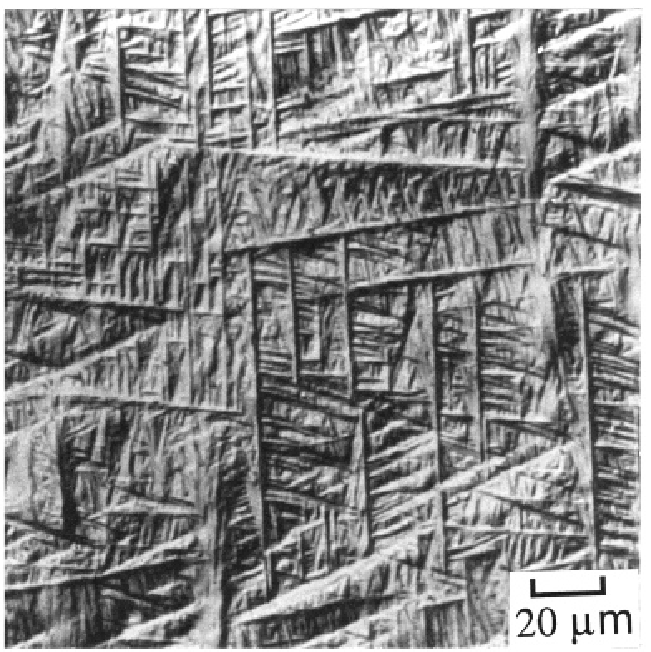}}\\
\subfigure(c){
\includegraphics[width=0.40\textwidth]{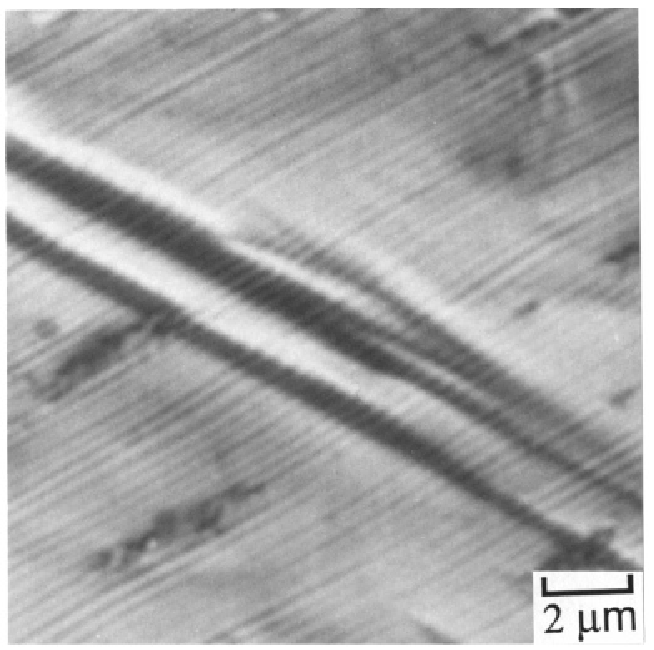}}
\caption{(a) Stylus record of surface upheavals caused by the bainite transformation. The horizontal graduations correspond to 1/100 of an inch, and the vertical magnification is ten times greater \cite{Ko:1952}. (b,c) Nomarski interference micrographs showing at a greater resolution, the displacements caused on a pre-polished sample of austenite by the formation of bainite \cite{Bhadeshia:1980}.}
\label{fig:Ko}
\end{figure}

The very first observation of the shape deformation due to clusters of bainite plates was by Ko and Cottrell \cite{Ko:1952}; a pre-polished sample of austenite was transformed into bainite and the resulting upheavals at the surface measured by traversing a stylus across the surface (Fig.~\ref{fig:Ko}a). The horizontal resolution is of the order of a few micrometers, which compares with a typical plate thickness of 0.25\,$\upmu$m. The work nevertheless had a profound effect on the development of the subject because it identified the first evidence for the disciplined movement of atoms involved in the growth of bainite.

Speich \cite{Speich:1958} subsequently used the surface relief to follow the growth kinetics of clusters of bainite plates using hot-stage optical microscopy, but did not report quantitative data on the nature of the relief. Somewhat higher resolution data obtained using Nomarski interference optical microscopy are shown in Fig.~\ref{fig:Ko}b,c.

\subsection{\bf Quantitative data}

None of the above measurements gave a quantitative value to the shape deformation of bainite. This is because light microscopy does not have the resolution to reveal the displacements due to individual platelets (the so-called sub-units of the transformation, \cite{Hehemann:1970}). Measurements of the shear strain using scratch displacements have the same resolution problem. The phases separating the bainitic ferrite sub-units have the effect of reducing the overall shear since they are benign during transformation. Srinivasan and Wayman realised that the shear strain data they measured using scratch displacements (Table~\ref{tab:Srinivasan}) are likely therefore to be underestimates of the actual shear; comparison with crystallographic theory indicated that the actual shear strain should be about twice that measured.

\begin{table}[htdp]
\caption{Magnitude of the shear strain averaged over a collection of bainite platelets containing also undeformed austenite \cite{Srinivasan:1968c}.}
\begin{center}
\begin{tabular}{lc}\midrule
Angle of shape shear & Shear strain $s$ \\[3pt]
\midrule
7$^\circ$9' & 0.1254 \\
8$^\circ$27' & 0.1254 \\
7$^\circ$0' & 0.1254 \\
7$^\circ$55' & 0.1254 \\
8$^\circ$20' & 0.1254 \\
5$^\circ$13' & 0.1254 \\
8$^\circ$24' & 0.1254 \\
6$^\circ$ & 0.1254 \\ \midrule
\end{tabular}
\end{center}
\label{tab:Srinivasan}
\end{table}%

\begin{figure}[!ht]
\begin{center}
\includegraphics[width=0.45\textwidth]{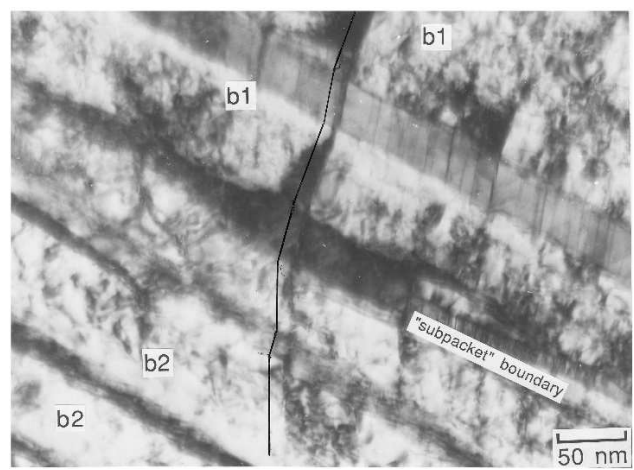}
\caption{Displacement of twin boundaries caused by individual sub-units of bainite.The ferrite variants `b1' and `b2' belong to separate sheaves \cite{Sandvik:1982}.}
\label{fig:Sandvik}
\end{center}
\end{figure}

An ingenious observation by Sandvik confirmed that the true shear is $s\approx0.22$; the measurement relied on the deflection caused by the shape deformation of bainite, of twins present in the austenite, and observed using thin-foil transmission electron microscopy (Fig.~\ref{fig:Sandvik}).

\begin{figure}[!ht]
\begin{center}
\includegraphics[width=0.4\textwidth]{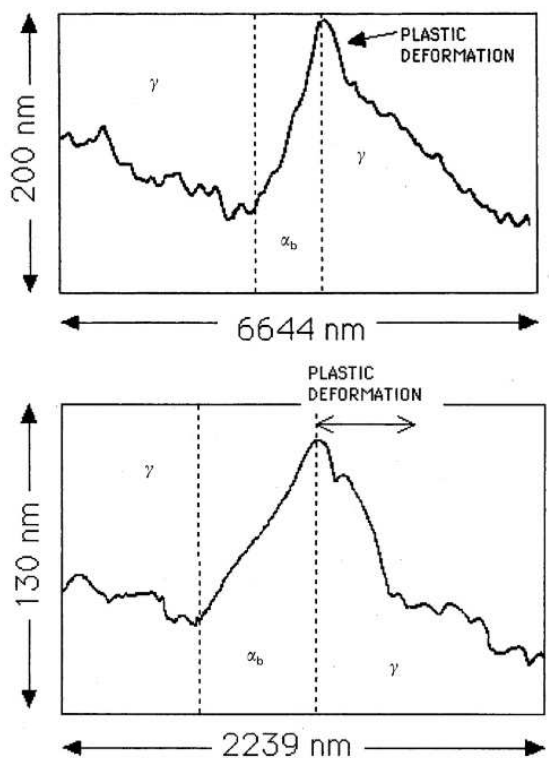}
\caption{Atomic force microscope scans across individual bainite sub-units showing the plastic deformation in the austenite ($\gamma$) adjacent to the bainitic ferrite $(\alpha_b$)  \cite{Swallow:1996}.}
\label{fig:AFM1}
\end{center}
\end{figure}

Atomic force microscopy (AFM) has the ability to resolve the displacements due to individual sub-units (Fig.~\ref{fig:displacements}b) and has provided considerable detail on the plastic accommodation effects caused in the adjacent austenite. Fig.~\ref{fig:AFM1} shows the plastic relaxation of the austenite adjacent to the bainitic ferrite. As emphasised earlier, it is important to measure the displacements due to individual platelets rather than that of a cluster of sub-units with intervening phases.

\begin{figure}[!ht]
\begin{center}
\includegraphics[width=0.4\textwidth]{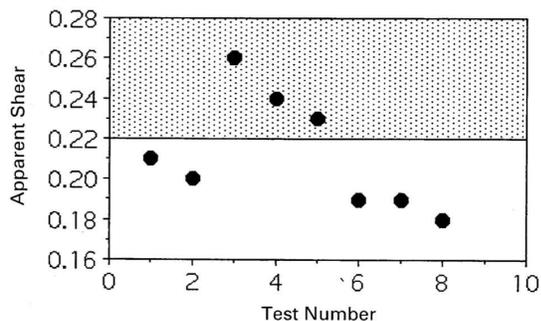}
\caption{AFM measurements of the apparent shear caused by the growth of bainite \cite{Swallow:1996}. The largest apparent shear is the minimum value of the true shear. }
\label{fig:AFM3}
\end{center}
\end{figure}

\begin{figure}[!ht]
\begin{center}
\includegraphics[width=0.4\textwidth]{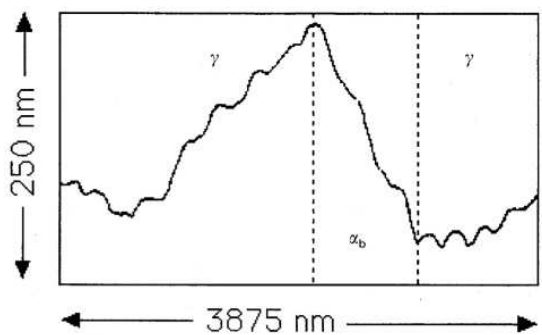}
\caption{An AFM scan showing an apparent shear strain of 0.26 caused by the growth of a bainite sub-unit \cite{Swallow:1996}.}
\label{fig:AFM2}
\end{center}
\end{figure}

AFM on its own does not give information about the orientation of the plate under the surface. Therefore, all measurements of the shear strain are {\em apparent}, dependent on the inclination of the plate and displacement direction relative to the free surface. Such data are illustrated in Fig.~\ref{fig:AFM3} for a number of measurements, with the shaded region representing reported values calculated using  the crystallographic theory of displacive transformations \cite{Wechsler:1953,BowlesMacKenzie:1954a,MacKenzie:1954,Bowles:1954c}. The largest recorded value of the apparent shear is 0.26, with the relevant displacements shown in Fig.~\ref{fig:AFM2}. This can be taken as the {\em minimum} value of the true shear, with $s\geq 0.26$. This is a large strain, far greater than a typical elastic strain of just $10^{-3}$, and its consequence, particularly on the associated strain energy (equation~\ref{eqn:christian}), should not be neglected in thermodynamic or kinetic assessments of the bainite transformation. \footnote{Reported scanning tunnelling microscopy data \cite{Yang:1996b,Fang:1996c,Bo:1998} on the surface relief caused by bainite are unfortunately of insufficient quality and lack quantitative interpretation to warrant detailed discussion.}

There has been much confusion on the microstructure of thermomechanically processed low-carbon (0.05\,wt\%) and high-niobium (0.1\,wt\%) pipeline steels, many millions of tonnes of which have been manufactured for the transmission of fossil fuels. It appears, however, that the confusion arises because insufficient parameters have been investigated in the published literature to enable the mechanisms to be resolved. A recent investigation has shown that the structure is correctly described as bainite \cite{Pei:2013b}. The surface relief data listed in Table~\ref{tab:s_a} are consistent with this, and with $s\geq 0.26$.

\begin{table}[htdp]
\caption{Apparent shear component $(s_a)$  of shape deformation due to the growth of bainite platelets, \cite{Pei:2013b}.}
\begin{center}
\begin{tabular}{cccc}
\midrule
Sample & Measured $s_a$ & Sample & Measured $s_a$\\
\midrule
1 & 0.18& 4 & 0.24\\
2 & 0.19& 5 & 0.24\\
3 & 0.24& 6 & 0.17\\
\midrule\end{tabular}
\end{center}
\label{tab:s_a}
\end{table}

 More recent work on nanostructured bainite that typically forms during transformation at 200$^\circ$C \cite{Peet:2011b}, has found that the shear strain associated with the slender plates of bainitic ferrite can be even larger at $s\geq 0.46$, \cite{Peet:2011b}, Table~\ref{tab:s_a}. This would, on the basis of equation~\ref{eqn:christian}, explain why the plates are only 20-40\,nm in thickness. However,  there is as yet no explanation of why the crystallography of the nanostructured bainite should be different from that of coarser bainite obtained at higher temperatures. 

\section{\bf Summary}
The fact that there is an invariant-plane shape change generated when austenite transforms into bainite is well established. It is also clear that the shear component of this shape deformation is large, with a minimum value of 0.26.  Such a shape deformation should not be neglected when considering the mechanisms of phase transformation, as is often done, even in modern literature \cite[e.g.]{Borgenstam:1996,Borgenstam:2009}. 

Perhaps it is more constructive to identify what remains to be done with respect to the displacive nature of the bainite reaction. The theoretical background to the crystallography of displacive reactions can be summarised as follows \cite{Wechsler:1953,BowlesMacKenzie:1954a,MacKenzie:1954,Bowles:1954c}:
$${\bf P_1 P_2 = RB}$$
where ${\bf B}$ is the Bain strain is the pure deformation that converts the crystal structure of austenite into ferrite, ${\bf R}$ is the rigid body rotation that in combination with ${\bf B}$ gives a total deformation that leaves a line invariant. The existence of such a line is a necessary condition for martensitic transformation \cite{Christian:2003,Bhadeshia:geometry}. The ${\bf RB}$ at the same time predicts the observed orientation relationship. ${\bf P}_1$ is the shape deformation matrix representing the invariant-plane strain and ${\bf P}_2$ is a lattice-invariant deformation that could be slip or twinning. The plane on which ${\bf P}_1$ occurs is of course the habit plane of the bainite in the present context. The point that emerges from this equation is that the shape deformation, orientation relationship and the crystallographic indices of the habit plane are all connected mathematically. All three of these quantities have been measured independently or as incomplete sets, but there has never been an experiment where they have been simultaneously determined for an individual bainite sub-unit. The reason of course is that this would be a difficult experiment given the fine scale of the sub-unit, but modern techniques including combination of atomic force microscopy, focused-ion beam machining and transmission electron microscopy could resolve the problem and bring closure to the subject.

The second issue highlighted here is the plastic accommodation of the shape deformation. Although the mechanical stabilisation of the austenite due to this effect is resolved quantitatively \cite{Chatterjee:2006}, it is not clear how the plastic accommodation influences the development of the crystallography of bainite.


\end{document}